\documentclass[11pt]{article}

\usepackage{graphicx}

\title{Simplified Variational Principles for Barotropic Fluid Dynamics}
\author{Asher Yahalom$^{a,b}$ and Donald Lynden-Bell$^{a,c}$ \\
$^a$ Institute of Astronomy, University of Cambridge\\
Madingley Road, Cambridge CB3 0HA, United Kingdom\\
$^b$ College of Judea and Samaria, Ariel 44284, Israel\\
$^c$ Clare College, University of Cambridge, Cambridge, United Kingdom\\
e-mail: dlb@ast.cam.ac.uk; asya@yosh.ac.il; }

\begin{document}
\maketitle

\newcommand{\beq} {\begin{equation}}
\newcommand{\enq} {\end{equation}}
\newcommand{\ber} {\begin {eqnarray}}
\newcommand{\enr} {\end {eqnarray}}
\newcommand{\eq} {equation}
\newcommand{\eqn} {equation }
\newcommand{\eqs} {equations }
\newcommand{\ens} {equations}
\newcommand{\mn}  {{\mu \nu}}
\newcommand {\er}[1] {equation (\ref{#1}) }
\newcommand {\ern}[1] {equation (\ref{#1})}
\newcommand {\ers}[1] {equations (\ref{#1})}
\newcommand {\Er}[1] {Equation (\ref{#1}) }

\begin {abstract}

We introduce a three independent functions variational formalism
for stationary and non-stationary barotropic flows.
This is less than the four variables which appear in the standard
equations of  fluid dynamics which are the velocity field $\vec v$
and the density $\rho$. It will be shown how in terms of our new
variable the Euler and continuity equations can be integrated in
the stationary case.

\end {abstract}
\noindent Keywords: Fluid dynamics, Variational principles
\\
\\
PACS number(s): 47.10.+g

\section {Introduction}

Variational principles for non-magnetic barotropic fluid dynamics
an Eulerian variational principle are well known. Initial attempts
to formulate Eulerian fluid dynamics in terms of a variational
principle, were described by Herivel \cite{Herivel}, Serrin
\cite{Serrin}, Lin \cite{Lin}. However, the variational principles
developed by the above authors were very cumbersome containing
quite a few "Lagrange multipliers" and "potentials". The range of
the total number of independent functions in the above
formulations ranges from eleven to seven which exceeds by many the
four functions appearing in the Eulerian and continuity equations
of a barotropic flow. And therefore did not have any practical use
or applications. Seliger \& Whitham \cite{Seliger} have developed
a variational formalism which can be shown to depend on only four
variables for barotropic flow. Lynden-Bell \& Katz
\cite{LynanKatz} have described a variational principle in
terms of two functions the load $\lambda$ (to be described below) and density $\rho$.
However, their formalism contains an implicit definition for the velocity $\vec v$
such that one is required to solve a partial differential equation in order
to obtain both $\vec v$ in terms of $\rho$ and $\lambda$ as well as its variations.
Much the same criticism holds for their general variational for non-barotropic flows \cite{KatzLyndeb}.
In this paper we overcome this limitation by paying the price of adding an additional
single function. Our formalism will allow arbitrary variations and the definition
of $\vec v$ will be explicit. Furthermore, we will show that for stationary flows
using somewhat different three variational
variables the Euler and continuity equations may be reduced to a single non-linear
algebraic equation for the density and thus can be integrated.

We anticipate applications for this study both for stability analysis of known fluid dynamics
 configurations and for designing efficient numerical schemes for integrating
the equations of fluid dynamics  \cite{Yahalom,YahalomPinhasi,YahPinhasKop,OphirYahPinhasKop}.

The plan of this paper is as follows:
We will review the basic equations of Eulerian fluid dynamics and give
a somewhat different derivation of Seliger \& Whitham's variational principle.
Then we will describe the three function variational principle for non-stationary
fluid dynamics. Finally we will give a different variational principle
for stationary fluid dynamics and we will show how in terms the new variational variables
the stationary equations of fluid dynamics that is the stationary Euler and continuity equations can be reduced
to a solution of a single non-linear algebraic equation for the density.

\section{Variational principle of non-stationary fluid dynamics}

Barotropic Eulerian fluids can be described in terms of four functions the
velocity $\vec v$ and density $\rho$. Those functions need to satisfy the
 continuity and Euler equations:
\beq
\frac{\partial{\rho}}{\partial t} + \vec \nabla \cdot (\rho \vec v ) = 0
\label{massconb}
\enq
\beq
\frac{d \vec v}{d t}=
\frac{\partial \vec v}{\partial t}+(\vec v \cdot \vec \nabla)\vec v  = -\frac{\vec \nabla p (\rho)}{\rho}
\label{Eulerb}
\enq
In which the pressure $p (\rho)$ is assumed to be a given function of the density.
Taking the curl of \ern{Eulerb} will lead to:
\beq
\frac{\partial{\vec \omega}}{\partial t} = \vec \nabla \times (\vec v \times \vec \omega)
\label{omegeq0}
\enq
in which:
\beq
\vec \omega = \vec \nabla \times \vec v
\label{vortic0}
\enq
is the vorticity. Equation (\ref{omegeq0}) describes the fact that the vorticity lines are "frozen"
within the Eulerian flow\footnote{The most general vortical flux and mass preserving flows that may
be attributed to vortex lines were found in \cite{LyndenB}}.

A very simple variational principle for non-stationary fluid dynamics
was described by Seliger \& Whitham \cite{Seliger} and is brought here mainly for completeness
using a slightly different derivation than the one appearing in the original paper.
This will serve as a starting point for the next section in which we
will show how the variational principle can be simplified further.
Consider the action:
\ber
A & \equiv & \int {\cal L} d^3 x dt
\nonumber \\
{\cal L} & \equiv & {\cal L}_1 + {\cal L}_2
\nonumber \\
{\cal L}_1 & \equiv & \rho (\frac{1}{2} \vec v^2 - \varepsilon (\rho)), \qquad
{\cal L}_2 \equiv  \nu [\frac{\partial{\rho}}{\partial t} + \vec \nabla \cdot (\rho \vec v )]
- \rho \alpha \frac{d \beta}{dt}
\label{Lagactionsimpb}
\enr
in which $\varepsilon (\rho)$ is the specific internal energy.
Obviously $\nu,\alpha$ are Lagrange multipliers which were inserted in such a
way that the variational principle will yield the following \ens:
\ber
& & \frac{\partial{\rho}}{\partial t} + \vec \nabla \cdot (\rho \vec v ) = 0
\nonumber \\
& & \rho \frac{d \beta}{dt} = 0
\label{lagmulb}
\enr
Provided $\rho$ is not null those are just the continuity \ern{massconb} and
the conditions that $\beta$ is comoving.
Let us take an arbitrary variational derivative of the above action with
respect to $\vec v$, this will result in:
\ber
\delta_{\vec v} A & = & \int d^3 x dt \rho \delta \vec v \cdot
[\vec v - \vec \nabla \nu - \alpha \vec \nabla \beta]
\nonumber \\
 & + & \oint d \vec S \cdot \delta \vec v \rho \nu
\label{delActionvb}
\enr
Provided that the above boundary term vanishes,
as in the case of astrophysical flows for which $\rho=0$ on the free flow
boundary, or the case in which the fluid is contained
in a vessel which induces a no flux boundary condition $\delta \vec v \cdot \hat n =0$
($\hat n$ is a unit vector normal to the boundary),
 $\vec v$ must have the following form:
\beq
\vec v = \hat {\vec v} \equiv \alpha \vec \nabla \beta + \vec \nabla \nu
\label{vformb}
\enq
this is nothing but Clebsch representation of the flow field (see for example \cite{Eckart}, \cite[page 248]{Lamb H.}).
Let us now take the variational derivative with respect to the density $\rho$, we obtain:
\ber
\delta_{\rho} A & = & \int d^3 x dt \delta \rho
[\frac{1}{2} \vec v^2 - w  - \frac{\partial{\nu}}{\partial t} -  \vec v \cdot \vec \nabla \nu]
\nonumber \\
 & + & \oint d \vec S \cdot \vec v \delta \rho  \nu + \int d^3 x \nu \delta \rho |^{t_1}_{t_0}
\label{delActionrhob}
\enr
in which $w = \frac{\partial(\rho \varepsilon )}{\partial \rho}$ is the specific enthalpy.
Hence provided that $\delta \rho$ vanishes on the boundary of the domain and in initial
and final times the following \eqn must be satisfied:
\beq
\frac{d \nu}{d t} = \frac{1}{2} \vec v^2 - w
\label{nueqb}
\enq
Finally we have to calculate the variation with respect to $\beta$
this will lead us to the following results:
\ber
\delta_{\beta} A & = & \int d^3 x dt \delta \beta
[\frac{\partial{(\rho \alpha)}}{\partial t} +  \vec \nabla \cdot (\rho \alpha \vec v)]
\nonumber \\
 & - & \oint d \vec S \cdot \vec v \rho \alpha \delta \beta
 - \int d^3 x \rho \alpha \delta \beta |^{t_1}_{t_0}
\label{delActionchib}
\enr
Hence choosing $\delta \beta$ in such a way that the temporal and
spatial boundary terms vanish in the above integral will lead to
the equation:
\beq
\frac{\partial{(\rho \alpha)}}{\partial t} +  \vec \nabla \cdot (\rho \alpha \vec v) =0
\enq
Using the continuity \ern{massconb} this will lead to the equation:
\beq
\rho \frac{d \alpha}{dt} = 0
\label{alphacon}
\enq
Hence for $\rho \neq 0$ both $\alpha$ and $\beta$ are comoving coordinates. Since
the vorticity can be easily calculated from \er{vformb} to be:
\beq
\vec \omega = \vec \nabla \times \vec v =  \vec \nabla \alpha \times \vec \nabla \beta
\label{vorticb}
\enq
Calculating $\frac{\partial{\vec \omega}}{\partial t}$ in which $\omega$ is
given by \ern{vorticb} and taking into
account both \ern{alphacon} and \ern{lagmulb} will yield \ern{omegeq0}.

\subsection{Euler's equations}
\label{Eulerequations}

We shall now show that a velocity field given by \ern{vformb}, such that the
functions $\alpha, \beta, \nu$ satisfy the corresponding equations
(\ref{lagmulb},\ref{nueqb},\ref{alphacon}) must satisfy Euler's equations.
Let us calculate the material derivative of $\vec v$:
\beq
\frac{d\vec v}{dt} = \frac{d\vec \nabla \nu}{dt}  + \frac{d\alpha}{dt} \vec \nabla \beta +
 \alpha \frac{d\vec \nabla \beta}{dt}
\label{dvform12b}
\enq
It can be easily shown that:
\ber
\frac{d\vec \nabla \nu}{dt} & = & \vec \nabla \frac{d \nu}{dt}- \vec \nabla v_k \frac{\partial \nu}{\partial x_k}
 = \vec \nabla (\frac{1}{2} \vec v^2 - w)- \vec \nabla v_k \frac{\partial \nu}{\partial x_k}
 \nonumber \\
 \frac{d\vec \nabla \beta}{dt} & = & \vec \nabla \frac{d \beta}{dt}- \vec \nabla v_k \frac{\partial \beta}{\partial x_k}
 = - \vec \nabla v_k \frac{\partial \beta}{\partial x_k}
  \label{dnablab}
\enr
In which $x_k$ is a Cartesian coordinate and a summation convention is assumed. Inserting the result from equations
(\ref{dnablab}) into \ern{dvform12b} yields:
\ber
\frac{d\vec v}{dt} &=& - \vec \nabla v_k (\frac{\partial \nu}{\partial x_k} +
 \alpha \frac{\partial \beta}{\partial x_k} ) + \vec \nabla (\frac{1}{2} \vec v^2 -w)
 \nonumber \\
&=& - \vec \nabla v_k v_k + \vec \nabla (\frac{1}{2} \vec v^2 - w)
 = - \frac{\vec \nabla p}{\rho}
\label{dvform2bb}
\enr
 This proves that the Euler equations can be derived from the action given in \ern{Lagactionsimpb} and hence
all the equations of  fluid dynamics can be derived from the above action
without restricting the variations in any way. Taking the curl of \ern{dvform2bb} will lead to \ern{omegeq0}.

\subsection{Simplified action}
\label{simpact}
The reader of this paper might argue that the authors have introduced unnecessary complications
to the theory of fluid dynamics by adding three  more functions $\alpha,\beta,\nu$ to the standard set
$\vec v,\rho$. In the following we will show that this is not so and the action given in \ern{Lagactionsimpb} in
a form suitable for a pedagogic presentation can indeed be simplified. It is easy to show
that the Lagrangian density appearing in \ern{Lagactionsimpb} can be written in the form:
\ber
{\cal L} & = & -\rho [\frac{\partial{\nu}}{\partial t} + \alpha \frac{\partial{\beta}}{\partial t}
+\varepsilon (\rho)] +
\frac{1}{2}\rho [(\vec v-\hat{\vec v})^2-\hat{\vec v}^2]
\nonumber \\
& + &  \frac{\partial{(\nu \rho)}}{\partial t} + \vec \nabla \cdot (\nu \rho \vec v )
\label{Lagactionsimpb4}
\enr
In which $\hat{\vec v}$ is a shorthand notation for $\vec \nabla \nu + \alpha \vec \nabla \beta $
(see \ern{vformb}). Thus ${\cal L}$ has three contributions:
\ber
{\cal L} & = & \hat {\cal L} + {\cal L}_{\vec v}+ {\cal L}_{boundary}
\nonumber \\
\hat {\cal L} &\equiv & -\rho [\frac{\partial{\nu}}{\partial t} + \alpha \frac{\partial{\beta}}{\partial t}
+\varepsilon (\rho)+\frac{1}{2}(\vec \nabla \nu + \alpha \vec \nabla \beta )^2]
\nonumber \\
{\cal L}_{\vec v} &\equiv & \frac{1}{2}\rho (\vec v-\hat{\vec v})^2
\nonumber \\
{\cal L}_{boundary} &\equiv & \frac{\partial{(\nu \rho)}}{\partial t} + \vec \nabla \cdot (\nu \rho \vec v )
\label{Lagactionsimp5b}
\enr
The only term containing $\vec v$ is ${\cal L}_{\vec v}$, it can easily be seen that
this term will lead, after we nullify the variational derivative, to \ern{vformb} but will otherwise
have no contribution to other variational derivatives. Notice that the term ${\cal L}_{boundary}$
contains only complete partial derivatives and thus can not contribute to the equations although
it can change the boundary conditions. Hence we see that \ers{lagmulb}, \ern{nueqb} and \ern{alphacon}
can be derived using the Lagrangian density $\hat {\cal L}$ in which $\hat{\vec v}$ replaces
$\vec v$ in the relevant equations. Furthermore, after integrating the four \eqs
(\ref{lagmulb},\ref{nueqb},\ref{alphacon}) we can insert the potentials $\alpha,\beta,\nu$
into \ern{vformb} to obtain the physical velocity $\vec v$.
Hence, the general barotropic fluid dynamics problem is changed such that instead of
solving the four equations
(\ref{massconb},\ref{Eulerb}) we need to solve an alternative set which can be
derived from the Lagrangian density $\hat {\cal L}$.

\subsection{The inverse problem}
\label{inverse}

In the previous subsection we have shown that given a set of functions $\alpha,\beta,\nu$
satisfying the set of equations described in the previous subsections,
one can insert those functions into \ern{vformb} and \ern{vorticb} to obtain the physical
velocity $\vec v$ and vorticity $\vec \omega$. In this subsection we will address the inverse problem
that is, suppose we are given the quantities $\vec v$ and $\rho$ how can one calculate the
potentials $\alpha,\beta,\nu$? The treatment in this section will follow closely (with minor changes)
the discussion given by Lynden-Bell \& Katz \cite{LynanKatz} and is given here for completeness.

\begin{figure}
\vspace{8cm}
\includegraphics{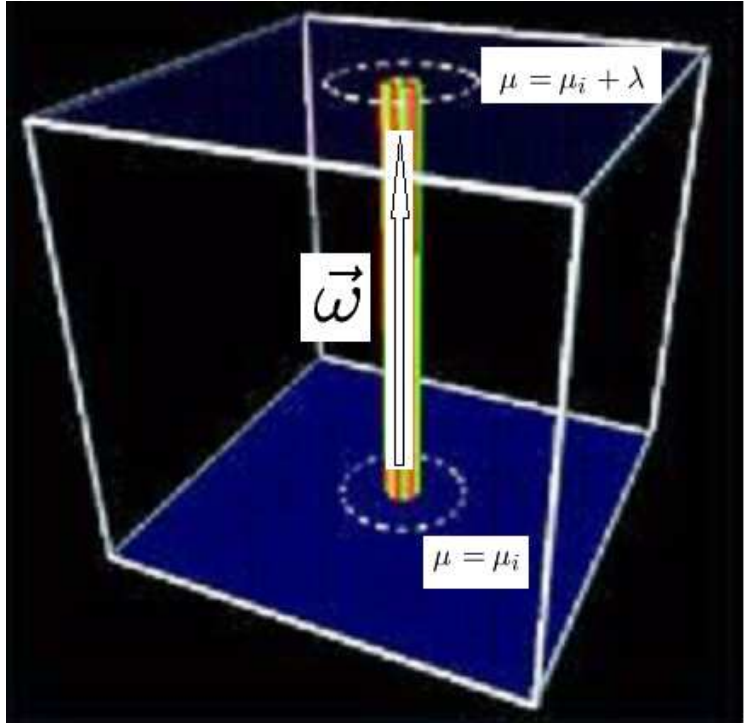}
\caption {A thin tube surrounding a vortex line}
\label{load}
\end{figure}

\begin{figure}
\vspace{8cm}
\includegraphics{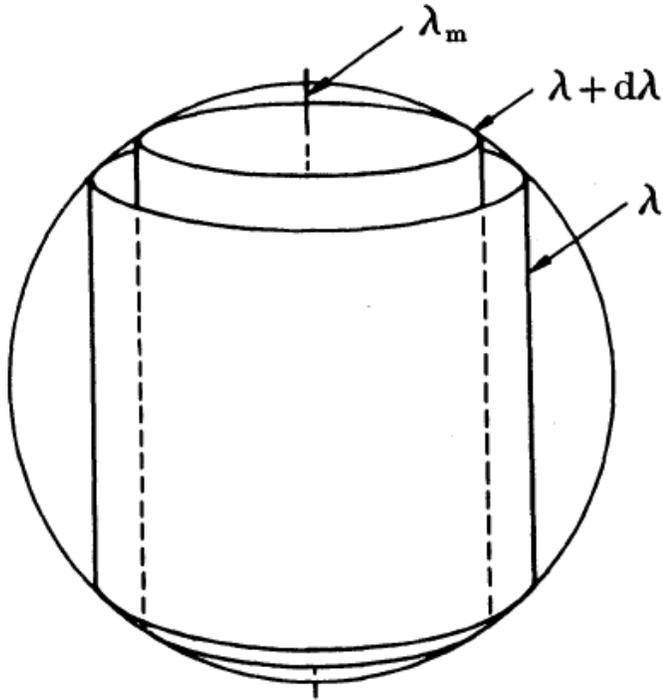}
\caption {Surfaces of constant load}
\label{loadsurface}
\end{figure}

Consider a thin tube surrounding a vortex line as described in figure \ref{load},
the vorticity flux contained within the tube which is equal to the circulation around the tube is:
\beq
\Delta \Phi = \int \vec \omega \cdot d \vec S = \oint \vec v \cdot d \vec r
\label{flux}
\enq
and the mass contained with the tube is:
\beq
\Delta M = \int \rho d\vec l \cdot d \vec S
\label{Mass}
\enq
in which $dl$ is a length element along the tube.
Since the vortex lines move with the flow by virtue of \ern{omegeq0} both the quantities $\Delta \Phi$
and $\Delta M$ are conserved and since the tube is thin we may define the conserved
load:
\beq
\lambda = \frac{\Delta M}{\Delta \Phi} = \oint \frac{\rho}{\omega}dl
\label{Load}
\enq
in which the above integral is performed along the field line.
Obviously the parts of the line which go out of the flow to regions
in which $\rho=0$ has a null contribution to the integral.
Since $\lambda$ is conserved is satisfies the equation:
\beq
 \frac{d \lambda }{d t} = 0.
\label{Loadcon}
\enq
By construction surfaces of constant load move with the flow and contain
vortex lines. Hence the gradient to such surfaces must be orthogonal to
the field line:
\beq
\vec \nabla \lambda \cdot \vec \omega = 0
\label{Loadortho}
\enq
Now consider an arbitrary comoving point on the vortex line and donate it by $i$,
and consider an additional comoving point on the vortex line and donate it by $r$.
The integral:
\beq
\mu(r)  = \int_i^r \frac{\rho}{\omega}dl + \mu(i)
\label{metage}
\enq
is also a conserved quantity which we may denote following Lynden-Bell \& Katz \cite{LynanKatz}
as the generalized metage. $\mu(i)$ is an arbitrary number which can be chosen differently for each
vortex line. By construction:
\beq
 \frac{d \mu }{d t} = 0.
\label{metagecon}
\enq
Also it is easy to see that by differentiating along the vortex line we obtain:
\beq
 \vec \nabla \mu \cdot \vec \omega = \rho
\label{metageeq}
\enq
At this point we have two comoving coordinates of flow, namely $\lambda,\mu$ obviously in a
three dimensional flow we also have a third coordinate. However, before defining the third coordinate
we will find it useful to work not directly with $\lambda$ but with a function of $\lambda$.
Now consider the vortical flux $\Phi(\lambda)$ within a surface of constant load  as described in figure \ref{loadsurface}
(the figure was given by Lynden-Bell \& Katz \cite{LynanKatz}). The flux is a conserved quantity
and depends only on the load $\lambda$ of the surrounding surface. Now we define the quantity:
\beq
 \alpha = \frac{\Phi(\lambda)}{2\pi}= \frac{C(\lambda)}{2\pi}
\label{chidef}
\enq
$C(\lambda)$ is the circulation along lines on this surface.
Obviously $\alpha$ satisfies the equations:
\beq
\frac{d \alpha}{d t} = 0, \qquad \vec \omega \cdot \vec \nabla \alpha = 0
\label{chieq}
\enq
Let us now define an additional comoving coordinate $\beta^{*}$
since $\vec \nabla \mu$ is not orthogonal to the $\vec \omega$ lines we can choose $\vec \nabla \beta^{*}$ to be
orthogonal to the $\vec \omega$ lines and not be in the direction of the $\vec \nabla \alpha$ lines,
that is we choose $\beta^{*}$
not to depend only on $\alpha$. Since both $\vec \nabla \beta^{*}$ and $\vec \nabla \alpha$
are orthogonal to $\vec \omega$,
$\vec \omega$ must take the form:
\beq
\vec \omega = A \vec \nabla \alpha \times \vec \nabla \beta^{*}
\enq
However, using \ern{vortic0} we have:
\beq
\vec \nabla \cdot \vec \omega = \vec \nabla A \cdot (\vec \nabla \alpha \times \vec \nabla \beta^{*})=0
\enq
Which implies that $A$ is a function of $\alpha,\beta^{*}$. Now we can define a new comoving function
$\beta$ such that:
\beq
\beta = \int_0^{\beta^{*}}A(\alpha,\beta^{'*})d\beta^{'*}, \qquad \frac{d \beta}{d t} = 0
\enq
In terms of this function we recover the representation given in \ern{vorticb}:
\beq
\vec \omega = \vec \nabla \alpha \times \vec \nabla \beta
\label{Bsakurai3}
\enq
Hence we have shown how $\alpha,\beta$ can be constructed for a known $\vec v,\rho$.
Notice however, that $\beta$ is defined in a non unique way since one can redefine
$\beta$ for example by performing the following transformation: $\beta \rightarrow \beta + f(\alpha)$
in which $f(\alpha)$ is an arbitrary function.
The comoving coordinates $\alpha,\beta$ serve as labels of the vortex lines.
Moreover the vortical flux can be calculated as:
\beq
\Phi = \int \vec \omega \cdot d \vec S = \int d \alpha d \beta
\enq

Finally we can use \ern{vformb} to derive the function $\nu$ for any point $s$ within the flow:
\beq
\nu(s) = \int_i^s (\vec v - \alpha  \vec \nabla \beta)\cdot d \vec r + \nu(i)
\enq
in which $i$ is any arbitrary point within the flow, the result will not depend on the trajectory taken
in the case that $\nu$ is single valued. If $\nu$ is not single valued on should introduce a cut,
 which the integration trajectory should not cross.

\subsection{Stationary fluid dynamics}
\label{Statflow}

Stationary flows are a unique phenomena of Eulerian fluid dynamics which has
no counter part in Lagrangian fluid dynamics. The stationary flow is defined
by the fact that the physical fields $\vec v,\rho$ do not depend on the
temporal coordinate. This however does not imply that the stationary potentials
$\alpha,\beta,\nu$ are all functions of spatial coordinates alone.
Moreover, it can be shown that choosing the potentials in such a way will lead
to erroneous results in the sense that the stationary equations of motion can
not be derived from the Lagrangian density $\hat {\cal L}$ given in \ern{Lagactionsimp5b}.
However, this problem can be amended easily
as follows. Let us choose $\alpha,\nu$ to depend on the spatial coordinates alone.
Let us choose $\beta$ such that:
\beq
\beta = \bar \beta - t
\label{betastas}
\enq
in which $\bar \beta$ is a function of the spatial coordinates. The Lagrangian
density $\hat {\cal L}$ given in \ern{Lagactionsimp5b} will take the form:
\beq
\hat {\cal L} = \rho \left(\alpha -\varepsilon (\rho)-
\frac{1}{2}(\vec \nabla \nu + \alpha \vec \nabla \beta)^2\right)
\label{stathatLb}
\enq
Varying the Lagrangian $\hat {L} = \int \hat {\cal L} d^3x$ with respect to $\alpha,\beta,\nu,\rho$
leads to the following equations:
\ber
& &  \vec \nabla \cdot (\rho \hat {\vec v} ) = 0
\nonumber \\
& & \rho \hat {\vec v} \cdot \vec \nabla \alpha = 0
\nonumber \\
& & \rho (\hat {\vec v} \cdot \vec \nabla \bar \beta - 1) = 0
\nonumber \\
& & \alpha= \frac{1}{2} \hat {\vec v}^2 + h
\label{statlagmulb}
\enr
$\alpha$ is thus the Bernoulli constant (this was also noticed in \cite{Yahalom}).
Calculations similar to the ones done in previous subsections will show that those equations
lead to the stationary Euler equations:
\beq
\rho (\hat {\vec v} \cdot \vec \nabla) \hat {\vec v} = -\vec \nabla p (\rho)
\label{Eulerstatb}
\enq

\section{A simpler variational principle of non-stationary fluid dynamics}

Lynden-Bell \& Katz \cite{LynanKatz} have shown that an Eulerian variational principle for
non-stationary fluid dynamics can be given in terms of two functions the density $\rho$ and
the non-magnetic load $\lambda$ defined in \ern{Load}. However, their velocity was given an
implicit definition in terms of a partial differential equation and its variations was constrained
to satisfy this equation. In this section we will propose a three function variational principle
in which the variations of the functions are not constrained in any way, part of our derivation
will overlap the formalism of Lynden-Bell \& Katz. The three variables will include the
density $\rho$, the non-magnetic load $\lambda$ and an additional function to be defined in the
next subsection. This variational principle is simpler than the Seliger \& Whitham variational
principle \cite{Seliger} which is given in terms of four functions and is more convenient
than the Lynden-Bell \& Katz \cite{LynanKatz} variational principle since the variations are not
constrained.

\subsection{Velocity representation}

Consider \ern{Loadortho}, since $\vec \omega$ is orthogonal to $\vec \nabla \lambda$
we can write:
\beq
\vec \omega = \vec K \times \vec \nabla \lambda
\enq
in which $\vec K $ is some arbitrary vector field. However, since $\vec \nabla \cdot \vec \omega=0$
it follows that $\vec K = \vec \nabla \theta $ for some scalar function theta. Hence we can write:
\beq
\vec \nabla \times \vec v= \vec \omega = \vec \nabla \theta \times \vec \nabla \lambda
\label{omthetlam}
\enq
This will lead to:
\beq
 \vec v =  \theta  \vec \nabla \lambda + \vec \nabla \nu
\label{vtheta}
\enq
For the time being $\nu$ is an arbitrary scalar function, the choice of notation will be justified
later. Consider now \ern{Loadcon}, inserting into this equation $\vec v$ given in \ern{vtheta} will result in:
\beq
\frac{d\lambda}{dt}  = \frac{\partial \lambda}{\partial t} + \vec v \cdot \vec \nabla \lambda=
\frac{\partial \lambda}{\partial t} + (\theta  \vec \nabla \lambda + \vec \nabla \nu) \cdot \vec \nabla \lambda=
0.
\label{loadbcon2}
\enq
This can be solved for $\theta$, the solution obtained is:
\beq
\theta= - \left(\frac{\frac{\partial \lambda}{\partial t} + \vec \nabla \lambda \cdot \vec \nabla \nu}
{|\vec \nabla \lambda|^2}\right)
\label{thetasol}
\enq
Inserting the above expression for $\theta$ into \ern{vtheta} will yield:
\beq
 \vec v = -\frac{\frac{\partial \lambda}{\partial t}}{|\vec \nabla \lambda|} \hat \lambda
+ \vec \nabla \nu - \hat \lambda (\hat \lambda \cdot \vec \nabla \nu)
\equiv -\frac{\frac{\partial \lambda}{\partial t}}{|\vec \nabla \lambda|} \hat \lambda
+ \vec \nabla^{*} \nu
\label{vtheta2}
\enq
in which $\hat \lambda =\frac{\vec \nabla \lambda}{|\vec \nabla \lambda|}$ is a unit vector perpendicular
to the load surfaces and $\vec \nabla^{*} \nu =\vec \nabla \nu - \hat \lambda (\hat \lambda \cdot \vec \nabla \nu)$
is the component of $\vec \nabla \nu$ parallel to the load surfaces. Notice
that the vector $\vec v - \vec \nabla \nu$ is orthogonal to the load surfaces and
that:
\beq
|\vec v - \vec \nabla \nu| = (\vec v - \vec \nabla \nu) \cdot \hat \lambda = \theta |\vec \nabla \lambda|
\Rightarrow \theta =\frac{(\vec v - \vec \nabla \nu) \cdot \hat \lambda}{|\vec \nabla \lambda|}
\label{thetavnu}
\enq
Further more by construction the velocity field $\vec v$ given by \ern{vtheta2} ensures that the load
surfaces are comoving. Let us calculate the circulation along $\lambda$ surfaces:
\beq
C(\lambda) = \oint_\lambda \vec v \cdot d \vec r = \oint_\lambda \vec \nabla^{*} \nu \cdot d \vec r
= \oint_\lambda \vec \nabla \nu \cdot d \vec r = [\nu]_\lambda
\enq
$[\nu]_\lambda$ is the discontinuity of $\nu$ across a cut which is introduced on the $\lambda$
surface. Hence in order that circulation $C(\lambda)$ on the load surfaces (and hence everywhere) will
not vanish $\nu$ must be multiple-valued. Following Lamb \cite[page 180, article 132, equation 1]{Lamb H.}
we write $\nu$ in the form:
\beq
\nu = C(\lambda) \bar{\nu}, \qquad [\bar{\nu}]_\lambda = 1
\label{nubar}
\enq
in terms of $\bar{\nu}$ the velocity is given as:
\beq
 \vec v =  -\frac{\frac{\partial \lambda}{\partial t}}{|\vec \nabla \lambda|} \hat \lambda
+ C(\lambda) \vec \nabla^{*} \bar{\nu}
\label{vtheta3}
\enq
And the explicit dependence of the velocity field $\vec v$ on the circulation
along the load surfaces $C(\lambda)$ is evident.

\subsection{The variational principle}

Consider the action:
\ber
A & \equiv & \int {\cal L} d^3 x dt
\nonumber \\
{\cal L} & \equiv & {\cal L}_1 + {\cal L}_2
\nonumber \\
{\cal L}_1 & \equiv & \rho (\frac{1}{2} \vec v^2 - \varepsilon (\rho)), \qquad
{\cal L}_2 \equiv  \nu [\frac{\partial{\rho}}{\partial t} + \vec \nabla \cdot (\rho \vec v )]
\label{Lagactionsimpc}
\enr
In which $\vec v$ is defined by \ern{vtheta2}. $\nu$ is not a simple Lagrange multiplier
since $\vec v$ is dependent on $\nu$ through \ern{vtheta2}. Taking the variational
derivative of ${\cal L}$ with respect to $\nu$ will yield:
\beq
\delta_{\nu} {\cal L} = \delta \nu [\frac{\partial{\rho}}{\partial t} + \vec \nabla \cdot (\rho \vec v )]
+ \rho \vec v \cdot \delta_{\nu} \vec v + \nu \vec \nabla \cdot (\rho \delta_{\nu} \vec v )
\enq
This can be rewritten as:
\beq
\delta_{\nu} {\cal L} = \delta \nu [\frac{\partial{\rho}}{\partial t} + \vec \nabla \cdot (\rho \vec v )]
+ \rho (\vec v - \vec \nabla \nu) \cdot \delta_{\nu} \vec v +  \vec \nabla \cdot (\rho \nu \delta_{\nu} \vec v )
\enq
Now by virtue of \ern{vtheta2}:
\beq
\delta_{\nu} \vec v = \vec \nabla^{*} \delta \nu
\enq
which is parallel to the load surfaces, while from \ern{vtheta} we see that $\vec v - \vec \nabla \nu$
is orthogonal to the load surfaces. Hence, the scalar product of those vector must be null and we can write:
\beq
\delta_{\nu} {\cal L} = \delta \nu [\frac{\partial{\rho}}{\partial t} + \vec \nabla \cdot (\rho \vec v )]
+  \vec \nabla \cdot (\rho \nu \vec \nabla^{*} \delta \nu )
\enq
Thus the action variation can be written as:
\ber
\delta_{\nu} A & = & \int d^3 x dt \delta \nu [\frac{\partial{\rho}}{\partial t} + \vec \nabla \cdot (\rho \vec v )]
\nonumber \\
 & + & \oint d \vec S \cdot \rho \nu \vec \nabla^{*} \delta \nu
\label{delActionnuc}
\enr
This will yield the continuity equation using the standard variational procedure.
Notice that the surface should include also the "cut" since the $\nu$ function is
in general multi valued.
Let us now take the variational derivative with respect to the density $\rho$, we obtain:
\ber
\delta_{\rho} A & = & \int d^3 x dt \delta \rho
[\frac{1}{2} \vec v^2 - w  - \frac{\partial{\nu}}{\partial t} -  \vec v \cdot \vec \nabla \nu]
\nonumber \\
 & + & \oint d \vec S \cdot \vec v \delta \rho  \nu + \int d^3 x \nu \delta \rho |^{t_1}_{t_0}
\label{delActionrhoc}
\enr
Hence provided that $\delta \rho$ vanishes on the boundary of the domain and in initial
and final times the following \eqn must be satisfied:
\beq
\frac{d \nu}{d t} = \frac{1}{2} \vec v^2 - w
\label{nueqc}
\enq
This is the same equation as \ern{nueqb} and justifies the use of the symbol $\nu$ in \ern{vtheta}.
Finally we have to calculate the variation of the Lagrangian density with respect to $\lambda$
this will lead us to the following results:
\ber
\delta_{\lambda} {\cal L} & = & \rho \vec v  \cdot \delta_{\lambda} \vec v +
\nu \vec \nabla \cdot (\rho \delta_{\lambda} \vec v )=
\rho (\vec v -\vec \nabla \nu ) \cdot \delta_{\lambda} \vec v +
 \vec \nabla \cdot (\rho \nu \delta_{\lambda} \vec v )
\nonumber \\
 & = &  \rho \theta |\vec \nabla \lambda|
 (\hat \lambda \cdot \delta_{\lambda} \vec v) +  \vec \nabla \cdot (\rho \nu \delta_{\lambda} \vec v )
\label{dellagranlam}
\enr
in \ern{thetavnu} was used. Let us calculate $\delta_{\lambda} \vec v$, after some straightforward manipulations one
arrives at the result:
\beq
\delta_{\lambda} \vec v = -\frac{\hat \lambda}{|\vec \nabla \lambda|}
\left[\frac{\partial (\delta \lambda)}{\partial t} + \vec v \cdot \vec \nabla \delta \lambda \right]
+ \theta \vec \nabla^{*} \delta \lambda
\label{delvlam}
\enq
Inserting \ern{delvlam} into \ern{dellagranlam} and integrating by parts will yield:
\beq
\delta_{\lambda} {\cal L} =  \delta \lambda \left[\frac{\partial (\rho \theta)}{\partial t}
+ \vec \nabla \cdot (\rho \theta \vec v)\right] +
\vec \nabla \cdot [\rho (\delta_{\lambda} \vec v \nu - \theta \vec v \delta \lambda)]
- \frac{\partial (\rho \theta \delta \lambda)}{\partial t}
\label{dellagranlam2}
\enq
Hence the total variation of the action will become:
\ber
\delta_{\lambda} A & = & \int d^3 x dt \delta \lambda
\left[\frac{\partial (\rho \theta)}{\partial t}
+ \vec \nabla \cdot (\rho \theta \vec v)\right]
\nonumber \\
 & + & \oint d \vec S \cdot (\delta_{\lambda} \vec v \nu - \theta \vec v \delta \lambda) \rho  -
 \int d^3 x \rho \theta \delta \lambda |^{t_1}_{t_0}
\label{delActionlambda}
\enr
Hence choosing $\delta \lambda$ in such a way that the temporal and
spatial boundary terms vanish in the above integral will lead to
the equation:
\beq
\frac{\partial{(\rho \theta)}}{\partial t} +  \vec \nabla \cdot (\rho \theta \vec v) =0
\enq
Using the continuity \ern{massconb} will lead to the equation:
\beq
\rho \frac{d \theta}{dt} = 0
\label{thetacon}
\enq
Hence for $\rho \neq 0$ both $\lambda$ and $\theta$ are comoving.
Comparing \ern{vformb} to \ern{vtheta} we see that $\alpha$ is analogue to $\theta$ and $\beta$
is analogue to $\lambda$ and all those variables are comoving. Furthermore, the $\nu$ function
in \ern{vtheta} satisfies the same equation as the $\nu$ appearing in \ern{vformb}
which is \ern{nueqc}. It follows immediately without the need for any additional
calculations that $\vec v$ given in \ern{vtheta} satisfies Euler's \ers{Eulerb},
the proof for this is given in subsection \ref{Eulerequations} in which one should
replace $\alpha$ with $\theta$ and $\beta$ with $\lambda$.
Thus all the equations of  fluid dynamics can be derived from the action (\ref{Lagactionsimpc})
without restricting the variations in any way. The reader should notice an important difference between the current
and previous formalism. In the current formalism $\theta$ is a dependent variable defined by \ern{thetasol},
while in the previous formalism the analogue quantity $\alpha$ was an independent variational
variable. Thus \er{thetacon} should be considered as some what complicated second-order partial differential
equation (in the temporal coordinate $t$) for $\lambda$ which should be solved simultaneously with
\ern{nueqc} and \ern{massconb}.

\subsection{Simplified action}

The Lagrangian density ${\cal L}$ given in \ern{Lagactionsimpc} can be written explicitly
in terms of the three variational variables $\rho,\lambda,\nu$ as follows:
\ber
{\cal L} & = & \hat {\cal L} + {\cal L}_{boundary}
\nonumber \\
\hat {\cal L} &\equiv & \rho \left[\frac{1}{2} \left(\frac{\frac{\partial{\lambda}}{\partial t} +
\vec \nabla \lambda \cdot \vec \nabla \nu}{|\vec \nabla \lambda|}\right)^2 - \frac{1}{2}(\vec \nabla \nu)^2 -
\frac{\partial \nu}{\partial t} - \varepsilon (\rho)\right]
\nonumber \\
{\cal L}_{boundary} &\equiv & \frac{\partial{(\nu \rho)}}{\partial t} + \vec \nabla \cdot (\nu \rho \vec v )
\label{Lagsimpc}
\enr
 Notice that the term ${\cal L}_{boundary}$
contains only complete partial derivatives and thus can not contribute to the equations although
it can change the boundary conditions. Hence we see that \ern{massconb}, \ern{nueqc} and \ern{thetacon}
can be derived using the Lagrangian density $\hat {\cal L}$ in which
$\vec v$ is given in terms of \ern{vtheta2} in the relevant equations. Furthermore, after integrating those
three equations we can insert the potentials $\lambda,\nu$
into \ern{vtheta2} to obtain the physical velocity $\vec v$.
Hence, the general barotropic fluid dynamics problem is altered such that instead of
solving the four equations (\ref{massconb},\ref{Eulerb}) we need to solve an alternative set of three equations which can be
derived from the Lagrangian density $\hat {\cal L}$. Notice that the specific choice of the
labelling of the $\lambda$ surfaces is not important in the above Lagrangian density one can replace:
$\lambda->\Lambda(\lambda)$, without changing the Lagrangian functional form. This means that only
the shape of the $\lambda$ surface is important not their labelling. In group theoretic language
this implies that the Lagrangian is invariant under an infinite symmetry group and hence should posses an infinite
number of constants of motion. In terms of the Lamb type function $\bar{\nu}$ defined in \ern{nubar},
the Lagrangian density given in \ern{Lagactionsimpc} can be rewritten in the form:
\ber
{\cal L} & = & \hat {\cal L} + {\cal L}_{boundary}
\nonumber \\
\hat {\cal L} &\equiv & \rho \left[\frac{1}{2} \left(\frac{\frac{\partial{\lambda}}{\partial t} +
C(\lambda)\vec \nabla \lambda \cdot \vec \nabla \bar{\nu}}{|\vec \nabla \lambda|}\right)^2 -
 \frac{1}{2}(C(\lambda)\vec \nabla \bar{\nu})^2 -
C(\lambda) \frac{\partial \bar{\nu}}{\partial t} - \varepsilon (\rho)\right]
\nonumber \\
{\cal L}_{boundary} &\equiv & \frac{\partial{(C(\lambda) \bar{\nu} \rho)}}{\partial t} +
\vec \nabla \cdot (C(\lambda) \bar{\nu} \rho \vec v )
\label{Lagactionsimpd}
\enr
Which emphasize the dependence of the Lagrangian on the the circulations
along the load surfaces $C(\lambda)$ which are given as initial conditions.

\subsection{Stationary fluid dynamics}

For stationary flows we assume that both the density $\rho$ and the load $\lambda$
are time independent. Hence the velocity field given in \ern{vtheta2} can be written as:
\beq
 \vec v = \vec \nabla \nu - \hat \lambda (\hat \lambda \cdot \vec \nabla \nu)
=  \vec \nabla^{*} \nu = C(\lambda)  \vec \nabla^{*} \bar{\nu}
\label{vtheta2stat}
\enq
thus the stationary flow is parallel to the load surfaces.
From the above equation we see that in the stationary case $\nu$ can be written in the form:
\beq
\nu = \nu_0 - f(\lambda,t)
\label{nu0}
\enq
in which $f(\lambda,t)$ is an arbitrary function and $\nu_0$ is independent of
the temporal coordinate.
Hence we can rewrite the velocity $\vec v$ as:
\beq
 \vec v =  \vec \nabla^{*} \nu_0 = C(\lambda)  \vec \nabla^{*} \bar{\nu}_0
\label{vtheta2statb}
\enq
Inserting \ern{nu0} and \ern{vtheta2statb} into \ern{nueqc}
will yield:
\beq
\frac{\partial f(\lambda,t)}{\partial t} = \frac{1}{2} \vec v^2 + w = B(\lambda)
\label{nueqstat}
\enq
in which $B(\lambda)$ is the Bernoulli constant. Integrating we obtain:
\beq
f(\lambda,t) = B(\lambda)t + g(\lambda)
\enq
the arbitrary $g(\lambda)$ function can be absorbed into $\nu_0$ and thus we rewrite \ern{nu0} in the form:
\beq
\nu = \nu_0 - B(\lambda)t
\label{nu0b}
\enq
Further more we can rewrite the conserved quantity $\theta$ given in \ern{thetasol}
as:
\beq
\theta= - \left(\frac{\vec \nabla \lambda \cdot \vec \nabla \nu}
{|\vec \nabla \lambda|^2}\right) = - \left(\frac{\hat{\lambda} \cdot \vec \nabla \nu_0}
{|\vec \nabla \lambda|}\right)+ t \frac{d B(\lambda)}{d\lambda}
\label{thetasolstat}
\enq
The Lagrangian density ${\cal L}$ given in \ern{Lagactionsimpc} can be written
in the stationary case taking into account \ern{vtheta2statb} and \ern{nu0b} as follows:
\ber
\hat {\cal L} &=& \rho \left[\frac{1}{2} ( \hat{\lambda} \cdot \vec \nabla \nu)^2
- \frac{1}{2}(\vec \nabla \nu)^2 +B(\lambda) - \varepsilon (\rho)\right]
\nonumber \\
&=& \rho \left[B(\lambda) - \frac{1}{2} (\vec \nabla^{*} \nu_0)^2 - \varepsilon (\rho)\right]
\nonumber \\
&=& \rho \left[B(\lambda) - \frac{1}{2} ( C(\lambda) \vec \nabla^{*} \bar{\nu}_0)^2 - \varepsilon (\rho)\right]
\label{Lagactionsimpcstat}
\enr
Taking the variational derivative of the Lagrangian density $\hat {\cal L}$ with respect to the mass
density $\rho$ will yield the Bernoulli \ern{nueqstat}. The variation of the
Lagrangian  $\hat L = \int d^3 x \hat {\cal L}$ with respect to $\nu_0$ will yield
the mass conservation equation:
\beq
\vec \nabla^{*}  \cdot( \rho \vec v) = 0
\label{mconlamb}
\enq
this form is equivalent to the standard stationary continuity equation \\
$\vec \nabla \cdot ( \rho \vec v) = 0$ since there is no mass flux orthogonal
to the load surfaces. Finally taking the variation of $\hat L$ with respect
to $\lambda$ will yield:
\beq
\rho \left[\frac{dB}{d\lambda} - \vec v \cdot \vec \nabla^{*} \left(\frac{\hat{\lambda} \cdot \vec \nabla \nu_0}
{|\vec \nabla \lambda|} \right)\right]= 0
\label{lambeq}
\enq
Which can be also obtained by inserting \ern{thetasolstat} into \ern{thetacon}.
Hence we obtained three equations (\ref{nueqstat},\ref{mconlamb},\ref{lambeq}) for the three spatial
functions $\rho, \lambda$ and $\nu_0$. Admittedly those equations do not have a particularly
simple form, we will obtain a somewhat better set of equations in the next section.

\section{Simplified variational principle for stationary fluid dynamics}

In the previous sections we have shown that fluid dynamics
can be described in terms of four first order differential equations and in
term of an action principle from which those equations can be derived.
An alternative derivation in terms of three differential equations one of
which (\ern{thetacon}) is second order has been introduced as well.
Those formalisms were shown to apply to both stationary and non-stationary fluid dynamics.
In the following  a different three functions formalism for stationary fluid dynamics is introduced.
In the suggested representation, the Euler and continuity equations can be integrated
leaving only an algebraic equation to solve.

Consider \ern{alphacon}, for a stationary flow it takes the form:
\beq
\vec v \cdot \vec \nabla \alpha = 0
\label{alphaconstat}
\enq
Hence $\vec v$ can take the form:
\beq
\vec v = \frac{\vec \nabla \alpha \times \vec K}{\rho}
\label{orthovb}
\enq
However, since the velocity field must satisfy the stationary mass conservation equation \ern{massconb}:
\beq
\vec \nabla \cdot (\rho \vec v ) = 0
\label{massconstatb}
\enq
We see that $\vec K$ must have the form $\vec K = \vec \nabla N$, where $N$ is an arbitrary function. Thus,
$\vec v$ takes the form:
\beq
\vec v = \frac{\vec \nabla \alpha \times \vec \nabla N}{\rho}
\label{orthov2b}
\enq
Let us now calculate $\vec v \times \vec \omega$ in which $\vec \omega$ is given by
\ern{vorticb}, hence:
\ber
\vec v \times \vec \omega &=& (\frac{\vec \nabla \alpha \times \vec \nabla N}{\rho}) \times
(\vec \nabla \alpha \times \vec \nabla \beta)
\nonumber \\
&=& \frac{1}{\rho} \vec \nabla \alpha (\vec \nabla \alpha \times \vec \nabla N) \cdot \vec \nabla \beta
\label{orthov3b}
\enr
Now since the flow is stationary $N$ can be at most a function of the three comoving coordinates
$\alpha,\bar \beta,\mu$ defined in subsections \ref{simpact} and \ref{Statflow}, hence:
\beq
\vec \nabla N  = \frac{\partial N}{\partial \alpha} \vec \nabla \alpha
+ \frac{\partial N}{\partial \bar \beta} \vec \nabla \bar \beta
+ \frac{\partial N}{\partial \mu} \vec \nabla \mu
\label{Ndivb}
\enq
Inserting \ern{Ndivb} into \ern{orthov3b} will yield:
\beq
\vec v \times \vec \omega =
\frac{1}{\rho} \vec \nabla \alpha \frac{\partial N}{\partial \mu}
(\vec \nabla \alpha \times \vec \nabla \mu) \cdot \vec \nabla \bar \beta
\label{orthov4b}
\enq
Rearranging terms and using vorticity formula (\ref{vorticb}) we can
simplify the above equation and obtain:
\beq
\vec v \times \vec \omega = -\frac{1}{\rho} \vec \nabla \alpha \frac{\partial N}{\partial \mu}
(\vec \nabla \mu \cdot \vec \omega)
\label{orthov5b}
\enq
However, using \ern{metageeq} this will simplify to the form:
\beq
\vec v \times \vec \omega = - \vec \nabla \alpha \frac{\partial N}{\partial \mu}
\label{orthov6b}
\enq
Now let us consider \ern{omegeq0}, for stationary flows this will take the form:
\beq
\vec \nabla \times (\vec v \times \vec \omega) = 0
\label{vortstatb}
\enq
Inserting \ern{orthov6b} into \ern{vortstatb} will lead to the equation:
\beq
\vec \nabla  (\frac{\partial N}{\partial \mu}) \times \vec \nabla \alpha = 0
\label{Beqstatimpb}
\enq
However, since $N$ is at most a function of $\alpha,\bar \beta,\mu$. It follows that
$\frac{\partial N}{\partial \mu}$ is some function of $\alpha$:
\beq
\frac{\partial N}{\partial \mu} = -F(\alpha)
\enq
This can be easily integrated to yield:
\beq
N = - \mu F(\alpha) + G(\alpha,\bar \beta)
\enq
Inserting this back into \ern{orthov2b} will yield:
\beq
\vec v = \frac{\vec \nabla \alpha \times (-F(\alpha) \vec \nabla \mu+
\frac{\partial G}{\partial {\bar \beta}} \vec \nabla \bar \beta) }{\rho}
\label{orthov7b}
\enq
Let us now replace the set of variables $\alpha,\bar \beta$ with a new set $\alpha',\bar \beta'$
such that:
\beq
\alpha' = \int F(\alpha) d\alpha, \qquad \bar \beta' = \frac{\bar \beta}{F(\alpha)}
\enq
This will not have any effect on the vorticity presentation given in \ern{vorticb} since:
\beq
\vec \omega = \vec \nabla \alpha \times \vec \nabla \beta = \vec \nabla \alpha \times \vec \nabla \bar \beta =
\vec \nabla \alpha' \times \vec \nabla \bar \beta'
\enq
However, the velocity will have a simpler presentation and will take the form:
\beq
\vec v = \frac{\vec \nabla \alpha' \times \vec \nabla(- \mu+ G'(\alpha',\bar \beta'))}{\rho}
\label{orthov8b}
\enq
in which $G'=\frac{G}{F}$. At this point one should remember that $\mu$ was defined in \ern{metage}
up to an arbitrary constant which can very between vortex lines. Since the lines
are labelled by their $\alpha',\bar \beta'$ values it follows that we can add an arbitrary function of
$\alpha',\bar \beta'$ to $\mu$ without effecting its properties. Hence we can define a new $\mu'$ such that:
\beq
\mu' =\mu-  G'(\alpha',\bar \beta')
\label{mupdefb}
\enq
Inserting \ern{mupdefb} into \ern{orthov8b} will lead to a simplified equation for $\vec v$:
\beq
\vec v = \frac{\vec \nabla \mu' \times \vec \nabla \alpha'}{\rho}
\label{orthov9b}
\enq
In the following the primes on $\alpha,\bar \beta,\mu$ will be ignored. It is obvious that
$\vec v$ satisfies the following set of equations:
\beq
\vec v \cdot \vec \nabla \mu = 0, \qquad  \vec v \cdot \vec \nabla \alpha = 0, \qquad
\vec v \cdot \vec \nabla \bar \beta = 1
\label{comovb}
\enq
to derive the right hand equation we have used both \ern{metageeq} and \ern{vorticb}.
Hence $\mu,\alpha$ are both comoving and stationary. As for $\bar \beta$ it satisfies \ern{statlagmulb}.

By vector multiplying $\vec v$ and $\vec \omega$ and using equations (\ref{orthov9b},\ref{vorticb})
we obtain:
\beq
\vec v \times \vec \omega = \vec \nabla \alpha
\label{vomsurfacesb}
\enq
this means that both $\vec v$ and $\vec \omega$ lie on $\alpha$ surfaces and provide a vector
basis for this two dimensional surface.

\subsection{The action principle}

In the previous subsection we have shown that if the velocity field $\vec v$
is given by \ern{orthov9b} than \ern{massconb}
is satisfied automatically for stationary flows. To complete the set of equations
we will show how the Euler \ers{Eulerb} can be derived from the Lagrangian:
\ber
L & \equiv & \int {\cal L} d^3 x
\nonumber \\
{\cal L} & \equiv & \rho (\frac{1}{2} \vec v^2 - \varepsilon (\rho))
\label{Lagactionsimpbstat}
\enr
In which $\vec v$ is given by \ern{orthov9b}
and the density $\rho$ is given by \ern{metageeq}:
\beq
\rho = \vec \nabla \mu \cdot \vec \omega = \vec \nabla \mu \cdot (\vec \nabla \alpha \times \vec \nabla \beta)
=\frac{\partial(\alpha,\beta,\mu)}{\partial(x,y,z)}
\label{metagecon2b}
\enq
In this case the Lagrangian density of \ern{Lagactionsimpbstat} will take the form:
\beq
{\cal L} = \rho \left(\frac{1}{2} (\frac{\vec \nabla \mu \times \vec \nabla \alpha}{\rho})^2 - \varepsilon (\rho)
\right)
\label{lagstatsimpb}
\enq
and can be seen explicitly to depend on only three functions.
The variational derivative of $L$ given in \ern{Lagactionsimpbstat} is:
\ber
\delta L & =& \int \delta {\cal L} d^3 x
\nonumber \\
\delta {\cal L} & = & \delta \rho ( \vec v^2 - w (\rho))+ \rho \vec v \cdot \delta \vec v
\label{varLag}
\enr
Let us make arbitrary small variations $\delta \alpha_i =(\delta \alpha,\delta \beta,\delta \mu)$ of the functions
$\alpha_i=(\alpha,\beta,\mu)$. Let us define the vector:
\beq
\vec \xi \equiv -\frac{\partial \vec r}{\partial \alpha_i} \delta \alpha_i
\label{xialdefb}
\enq
This will lead to the equation:
\beq
\delta \alpha_i = -\vec \nabla \alpha_i \cdot \vec \xi
\label{delalb}
\enq
 Making a variation of $\rho$ given in \ern{metagecon2b} with respect to $\alpha_i$ will yield:
\beq
\delta \rho = - \vec \nabla \cdot (\rho \vec \xi)
\label{delrho}
\enq
(for a proof see for example \cite{Katz}).
Calculating $\delta \vec v$ by varying \ern{orthov9b} will give:
\beq
\delta \vec v = -\frac{\delta \rho}{\rho} \vec v +
\frac{1}{\rho} \vec \nabla \times (\rho \vec \xi \times \vec v)
\label{delv2b}
\enq
Inserting \eqs (\ref{delrho},\ref{delv2b}) into \ern{varLag}
will yield:
\ber
\delta {\cal L} &=& \vec v \cdot \vec \nabla \times (\rho \vec \xi \times \vec v)
- \delta \rho (\frac{1}{2} \vec v^2 + w)
\nonumber \\
&=& \vec v \cdot \vec \nabla \times (\rho \vec \xi \times \vec v)
+ \vec \nabla \cdot (\rho \vec \xi) (\frac{1}{2} \vec v^2 + w )
\label{delcalLb}
\enr
Using the well known vector identity:
\beq
\vec A \cdot \vec \nabla \times (\vec C \times \vec A)=
\vec \nabla \cdot ((\vec C \times \vec A) \times \vec A)+
(\vec C \times \vec A) \cdot \vec \nabla \times \vec A
\label{veciden1}
\enq
and the theorem of Gauss we can write now \ern{varLag} in the form:
\ber
\delta L & = &
 \oint d \vec S \cdot [ (\vec \xi \times \vec v)\times \vec v
+(\frac{1}{2} \vec v^2 + w )\vec \xi]\rho
\nonumber \\
& + & \int d^3 x  \vec \xi \cdot [ \vec v \times \vec \omega-\vec \nabla (\frac{1}{2} \vec v^2 + w )]\rho
\label{delLagaction2simplb}
\enr
Suppose now that $\delta L = 0$ for a $\vec \xi$ such that the boundary term in the above equation
is null but that $\vec \xi$ is otherwise arbitrary, then it entails the equation:
\beq
\rho \vec v \times \vec \omega-\rho \vec \nabla (\frac{1}{2} \vec v^2 + w ) = 0
\label{Eulerstat22b}
\enq
Using the vector identity :
\beq
\frac{1}{2} \vec \nabla (\vec v^2) = (\vec v \cdot \vec \nabla) \vec v +
 \vec v \times (\vec \nabla \times \vec v)
\label{veciden2}
\enq
and rearranging terms we recover the stationary Euler equations:
\beq
\rho (\vec v \cdot \vec \nabla)\vec v = -\vec \nabla p
\label{Eulerstat2b}
\enq

\subsection{The integration of the stationary Euler equations}

Let us now combine \ern{Eulerstat22b} with \ern{vomsurfacesb},
this will yield:
\beq
\vec \nabla \alpha =  \vec \nabla (\frac{1}{2} \vec v^2 + w )
\label{Eulerstat22bb}
\enq
which can easily be integrated to yield:
\beq
 \alpha =  \frac{1}{2} \vec v^2 + w (\rho)
\label{Eulerstat22c}
\enq
in which an arbitrary constant is absorbed into $\alpha$, this is consistent with \ern{statlagmulb} and implies that
$\alpha$ is the Bernoulli constant. Let us now insert $\vec v$
given by \ern{orthov9b} into the above equation:
\beq
 \alpha =  \frac{1}{2} \left(\frac{\vec \nabla \mu \times \vec \nabla \alpha}{\rho}\right)^2 + w (\rho)
\label{Eulerstat22d}
\enq
This can be rewritten as:
\beq
w (\rho) \rho^2 -\rho^2 \alpha +\frac{1}{2} (\vec \nabla \mu \times \vec \nabla \alpha)^2 = 0
\label{Eulerstat22e}
\enq
By defining:
\ber
\Pi &\equiv& \rho^2
\nonumber \\
W(\Pi) &\equiv& w(\rho)
\enr
We obtain:
\beq
W(\Pi) \Pi - \Pi  \alpha + \frac{1}{2} (\vec \nabla \mu \times \vec \nabla \alpha)^2 = 0
\label{Eulerstat22f}
\enq
If one chooses an $\alpha$ and $\mu$ functions such that the above nonlinear
equation has a positive (or zero) solution for $\Pi$ at every point in space than
one obtains a solution of both the Euler and continuity equations in which
$\vec v$ is given by \ern{orthov9b} and $\rho=+\sqrt{\Pi}$. One should notice,
however, that $\alpha$ and $\mu$ are not arbitrary functions. $\alpha$ is a function
of the load defined in \ern{Load} and thus according to \ern{Loadortho} must satisfy the equation:
\beq
\vec \nabla \alpha \cdot \vec \omega = 0
\label{alphaortho}
\enq
that is both the velocity and vorticity fields must lie on the alpha surfaces. While
for the velocity $\vec v$ this is assured by \ern{orthov9b} this is not assured for
the vorticity. Furthermore, $\mu$ is not an arbitrary function it is a metage function
an thus must satisfy \ern{metageeq}. Moreover, obtaining a solution does
not imply that the obtained solution is stable.
Having said that we notice that the technique requires only the
solution of the algebraic \ern{Eulerstat22f} and does involve solving any differential
equations.

\section {Conclusion}

In this paper we have reviewed Eulerian variational principles for non-stationary barotropic fluid dynamics
and introduced a simpler three independent functions variational formalisms for stationary and non-stationary
barotropic flows.
This is less than the four variables which appear in the standard equations of  fluid dynamics which
are the velocity field $\vec v$ and the density $\rho$. We have shown how in terms
of our new variables the stationary Euler and continuity equations can be integrated,
such that the stationary Eulerian fluid dynamics problem is reduced to a non-linear
algebraic equation for the density.

The problem of stability analysis and the description of
numerical schemes using the described variational principles exceed the scope of this paper.
We suspect that for achieving this we will need to add additional
constants of motion constraints to the action as was done by \cite{Arnold1,Arnold2}
see also \cite{YahalomKatz}, hopefully this will be discussed in a future paper.

\end {document}